# SIGNIFICANCE OF THE TEAMWORK IN AGILE SOFTWARE ENGINEERING

**M. Rizwan Jameel Qureshi, Sohayp Abo Alshamat**
Faculty of Computing and Information Technology, King Abdulaziz University,
Jeddah, Kingdom of Saudi Arabia
anriz@hotmail.com, robbn84@hotmail.com

**Fatima Sabir**
Faculty of Computer Science and Information Technology,
Lahore Leads University
fatimasajidnazir@gmail.com
Cell # (+966-536474921)

*ABSTRACT: A Software Engineering project depends significantly on team performance, as does any activity that involves human interaction. In the last years, the traditional perspective on software development is changing and agile methods have received considerable attention. Among other attributes, the ageists claim that fostering creativity is one of the keys to response to common problems and challenges of software development today. The development of new software products requires the generation of novel and useful ideas. It is a conceptual framework introduced in the Agile Manifesto in 2001. This paper is written in support of agile practices in terms of significance of teamwork for the success of software projects. Survey is used as a research method to know the significance of teamwork.*

Key words: software engineering team, Creativity of teamwork, te*am*

## 1. INTRODUCTION

There are 3 reasons for people to work in teams, and why we require them in this class:

1) Research shows that people learn better when working with others in teams, as opposed to working in isolation or competing against each other. This is a very important observation, especially considering the expectation that all of us need to continue learning long after we leave school and teamwork provides a great vehicle for this purpose.
2) You need to learn and practice team and small group communication skills, both of which are important tools for pursuing a successful career in design and engineering.
3) The Accreditation Board for Engineering and Technology (ABET), that sets the standards engineering programs in the US, specified in its Engineering Criteria 2000, that engineering graduates should be able to function on multidisciplinary teams. The National Association for Industrial Technology (NAIT), that sets the standards for industrial technology programs in the US, also specifies team work as a necessary educational outcome.

Although many universities have recognized the need to assign group projects and have begun efforts to improve engineering and computer science curricula in this regard, students seldom receive any training on how to function collaboratively before such assignments are given, and little attention is given to how teams are formed. Consequently, teams often fail to function effectively. Furthermore, students do not learn much from participating on dysfunctional teams and often develop negative views about the value of teamwork.

The remainder of this paper is organized as follows: Section 2 describes research achievement concerned with teamwork. Section 3 covers the problem & proposes a solution of the current problem. Section 4 validates the proposed solution using survey technique. Finally, conclusion is given in the final section.

## 2. RELATED WORK

In this paper [1] we show that the success of a software project depends on human and social factors. But surprisingly, the human side of software development has been ignored. Knowledge management and creativity are crucial for software development. To highlight this important relationship we compared phases and roles of XP with phases and roles in creativity processes. Based on such comparison it is claimed that creativity can be made larger part of software development when performed in the form of XP. Creative work involves collaboration and interaction, which is also important, strives in agile development. The Extreme Programming methodology includes implicitly central aspects of a creative teamwork. These aspects can be organized according to the structure that the team adopts and the performance that characterizes to the team. The structure that the team adopts and specially the different roles that the methodology advises to define, nearly correspond with the roles at the interior of a creative team. The performance that characterizes the team through certain advisable practices, from the perspective of creativity, constitutes the necessary basic conditions, although nonsufficient, in order to favor the group creative performance. These conditions called practices in XP methodology are accompanied by concrete phases of constituent activities of an agile software development process, which is possible to correspond with the creative process, which is fundamental to the creative performance.

About this paper [2] introduce the agile method called Extreme Programming (XP) is analyzed and evaluated from the perspective of the creativity, in particular the creative performance and structure required at the teamwork level. The conclusion is that XP has characteristics that ensure the creative performance of the team members, but we believe that it can be fostered from a creativity perspective. And Author concludes The Extreme Programming methodology includes implicitly central aspects of a creative teamwork. These aspects can be organized according to the structure that the team adopts and the performance that characterizes to the team. The structure that the team adopts and specially the different roles that the methodology advises to define, nearly correspond with the roles at the interior of a creative team. The performance that characterizes the team through certain advisable practices, from the perspective of creativity, constitutes the necessary basic conditions, although nonsufficient, in order to favor the group creative performance.

[3] is written about relationship to the structure of the team and performance purposes of the team. What roles (or their functionality) of a creative team are included (or should be included) in the XP team? Is a goal of the XP team to generate novelty ideas? Is the performance (operating) of a XP team equivalent to the creative process (divergent thinking/convergent thinking)?

Work in [5] argues for the use of diagnosis and action planning to improve teamwork in agile software development. The action planning focused on improving shared leadership, team orientation and learning. The improvement project provided most new insight for the mature team. The innovation and development of new products is an interdisciplinary issue, they are interested in the study of the potential of new concepts and techniques to foster creativity in software engineering.

A case study is reported on the use of wikis in an undergraduate software engineering course at the IT Department at King Saud University [6]. The IT Department is a female only Department. The wiki was used by students for their software development project. The objective of this study is to investigate how the wiki tool can be used to support the development of students' teamwork skills. This study showed that the wiki played an important role in supporting teamwork. One of the important aspects of using wikis for software projects is that both students and the instructor can visualize the team development process throughout the project. For students, the wiki allows them to manage their team, communicate with other team members, monitor progress, and monitor team activity (who is participating and who is not). For the instructors, the wiki enables them to stay informed about student interactions and team status, thereby detecting problems early on and ensuring that all teams are successful.

The novel approach for assessing student teamwork performance has been present in this [7], with a specific case illustrating the use of the approach. The approach combines the previous research on cooperative learning, collaborative learning, team management and learning style theory. The approach is suitable for software engineering courses and can be tailored to specific courses or requirement in teamwork skill development. They have used the approach in several courses and the results are positive and promising. The approach assesses a student team performance in terms of the team performance and the contributions of the student to the team. This encourages students actively participate team activities through a series of team tasks. The approach in forming a team based on learning styles helps to maintain the learning style diversity and to balance the knowledge distribution within a team. Those features of the approach require students to be involved in teamwork in a more forceful way and hence hopefully improving their teamwork skills. The provision of teamwork guidelines and templates helps to reduce the difficulties and hardness in performing teamwork. The rather rigid calculation of the performance is mainly for fairness and possible automation tool support.

## 3. PROBLEM DEFINITION AND THE PROPOSED SOLUTION

The success of a software project depends on human and social factors. But surprisingly, the human side of software development has been ignored. Knowledge management and creativity are crucial (where team works are the basic component) for software development. To highlight this important relationship between team works in software engineering to success any project, we proposed this work. There is need to resolve the teamwork problems. This solution depends on three phases:

Goal 1) Members must work as a team.

Goal 2) A person who has high knowledge must improve the weaknesses of other team members.

Goal 3) All members think logically.

## 4. VALIDATION OF THE PROPOSED SOLUTION

Is that working as team work is much better than individual work especially for big projects? Validation of proposed solution is conducting a survey through the questioner. The following part is description of questionnaire result which purpose to approve our goals which are:

### A. CUMULATIVE STATISTICAL ANALYSIS OF GOAL 1.

PM must be created many ways from our experience according to structure of team to improve creativity. Dividing the responsibilities by create team not one person do everything and burden all problem. The result of the analysis of the Goal 1 is shown in Table 1.

Table 1 Cumulative Analysis of Goal 1

| Q. No | Str. Disagree | Disagree | Neutral | Agree | Str. Agree |
|---|---|---|---|---|---|
| 1 | - | 3 | 5 | 11 | 13 |
| 2 | 2 | 4 | 14 | 11 | 1 |
| 3 | 2 | 1 | 5 | 14 | 10 |
| 4 | - | 4 | 11 | 11 | 6 |
| Total | 4 | 12 | 35 | 47 | 30 |
| Avg. | 3.125 | 9.375 | 27.34 | 36.71 | 23.43 |

As it is clear from the cumulative descriptive analysis of goal 1 that 36% of the sample agreed that we can find the long term learning requirements with the proposed solution and 23.43 % strongly agreed to it.9.375% disagreed to it and 3.125% strongly disagreed to it while 27.34% remained neutral as shown below in "Fig. 1".

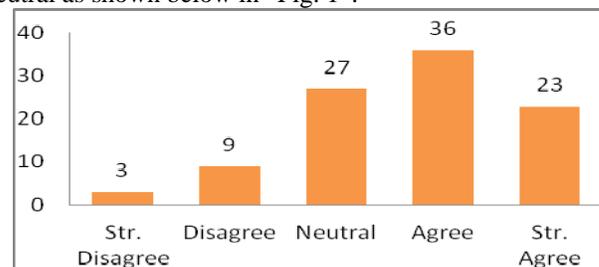

Figure 1 Graph showing the cumulative results of questionnaire for Goal 1

## *Cumulative Analysis of Goal 2*

A project manager must balance knowledge in his team members. Seniors must respect and help their juniors to transfer knowledge. The results are shown in Table 2.

Table 2 Cumulative Analysis of Goal 2

| Q. No | Str. Disagree | Disagree | Neutral | Agree | Str. Agree |
|---|---|---|---|---|---|
| 1 | - | 4 | 11 | 11 | 6 |
| 2 | - | 3 | 9 | 14 | 6 |
| 3 | - | 5 | 5 | 12 | 10 |
| 4 | 1 | 1 | 10 | 10 | 10 |
| Total | 1 | 13 | 35 | 47 | 32 |
| Avg. | 0.78125 | 10.15625 | 27.343 | 36.718 | 25 |

As it is clear from the cumulative descriptive analysis of goal 2 that 36.718% of the sample agreed that we can find the long term learning requirements with the proposed solution and 25 % strongly agreed to it.10.15% disagreed to it and 1% strongly disagreed to it while 27% remained neutral as shown below in "Fig. 2".

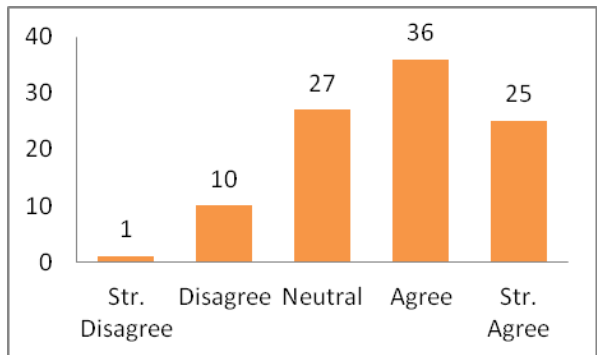

Figure 2 Graph showing the cumulative results of Questionnaire for Goal 2

### B. Cumulative Analysis of Goal 3

Not just produces and roles the employ of origination implement without thinking do in efficient way. The result of the analysis of the goal 3 is shown in Table 3.

Table 3 Cumulative Analysis of Goal 3

| Q. No | Str. Disagree | Disagree | Neutral | Agree | Str. Agree |
|---|---|---|---|---|---|
| 1 | 1 | 9 | 10 | 9 | 3 |
| 2 | 1 | 2 | 8 | 16 | 5 |
| 3 | 2 | 4 | 6 | 15 | 5 |
| Total | 4 | 15 | 24 | 40 | 13 |
| Avg. | 4.1667 | 15.625 | 4.166 | 25 | 41.66 |

As it is clear from the cumulative descriptive analysis of Goal 3 that 25% of the sample agreed that we can find the long term learning requirements with the proposed solution and 41.66 % strongly agreed to it, 15.6% disagreed to it, 34.16% strongly disagreed to it and 27.34% remained neutral as shown below in figure 3.

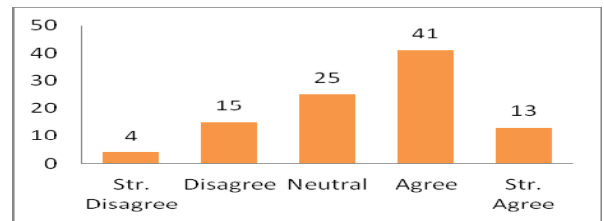

Figure 3 Graph showing the cumulative results of questionnaire for Goal 3

## 5. CONCLUSION

Creativity and innovation are essential skills in almost any teamwork. Having a team that can solve problems quickly and effectively with a little creative thinking is beneficial to everyone. The performance of a team depends not only on the competence of the team itself in doing its work, but also on the organizational context. The organizational conditions in which the team is inserted are very important too. If workers see that their ideas are encouraged and accepted, they will be more likely to be creative, leading to potential innovation in the workplace. The creation of a collaborative work environment will foster the communication between employees and reward those that work together to solve problems. Encouraging team members to take risks, the opposite of creativity is fear, and then it is necessary to create an environment that is free from fear of failure: failures are a learning tool. The Extreme Programming methodology includes implicitly central aspects of a creative teamwork. This work is supported with validation in which 59.03% people supported it while 14.41% disagreed to it as shown in figure 4.

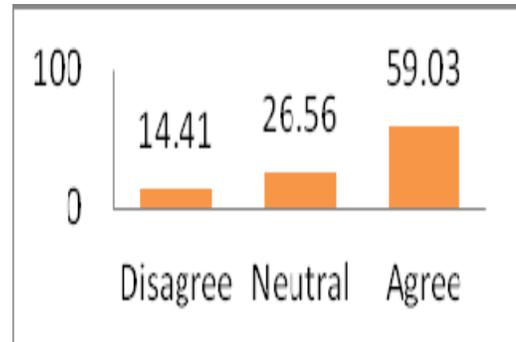

Figure 4 Final Results of the Proposed Solution